\documentclass{PoS}

\title{A More Complete Phenomenology of Tau Lepton Induced Air Showers}

\ShortTitle{Tau Lepton Showers}

\author{\speaker{A. Cummings} $^{1,2 \dag}$, R. Aloisio$^{1,2}$, M. Bertaina$^{3,4}$, F. Bisconti$^{3,4}$, F. Fenu$^{3,4}$, F. Salamida$^{2,5}$\\
\llap{$^1$}Gran Sasso Science Institute, L'Aquila, Italy\\
\llap{$^2$}INFN, Laboratori Nazionali Gran Sasso, Assergi (L'Aquila), Italy\\
\llap{$^3$}Universita di Torino, Torino, Italy\\
\llap{$^4$}INFN Torino, Torino, Italy\\
\llap{$^5$}Universita dell'Aquila, Dipartimento di Scienze Fisiche e Chimiche, L'Aquila, Italy\\
\llap{$^\dag$}E-mail: \email{austinlee.cummings@gssi.it}}

\abstract{Many proposed and upcoming experiments seek to observe signals from upward going air showers initiated by tau leptons resulting from neutrino interactions inside the Earth. To save calculation time, event estimations for these observation methods are usually performed while making several assumptions about the showers themselves, which simplifies their rich phenomenology and may or may not lead to inaccuracies in results. Here, we present results of extensive CORSIKA simulations of upward going tau initiated showers in the energy range 1~PeV to 10~EeV. Specifically, we monitor the Cherenkov emission, the charged particle distributions, and the timing of the showers for different geometric configurations. We analyze the impact of the decay length and different decay modes of the tau on particle distributions and compare to primaries usually utilized to simulate a tau shower, such as gammas, electrons, and protons. We also check the accuracy of many of the usual assumptions of these showers and analyze the often ignored muon channel of the tau decay.}

\FullConference{36th International Cosmic Ray Conference -ICRC2019-\\
		July 24th - August 1st, 2019\\
		Madison, WI, U.S.A.}

\begin{document}

\section{Introduction}
Tau neutrinos are produced in oscillations of cosmic neutrinos as they travel from their sources to Earth. These neutrinos should produce a flux of tau leptons after propagation through the Earth, where they undergo charged and neutral current interactions, as well as energy losses, and possible decay/regeneration. The exiting tau lepton can decay in the atmosphere, producing an upward moving extensive air shower, referred to as an UAS. UAS are interesting to neutrino astronomers, as they offer a method for probing large detection areas fairly easily, either by observing the side of thick mountains, or by observing Earth's surface from altitude. As of yet, there has not been a confirmed measurement of an astrophysical tau lepton, but the observation of a UAS would help confirm the current cosmic origin interpretation of the IceCube neutrino flux.

\begin{wrapfigure}{R}{0.5\textwidth}
\includegraphics[width=\linewidth]{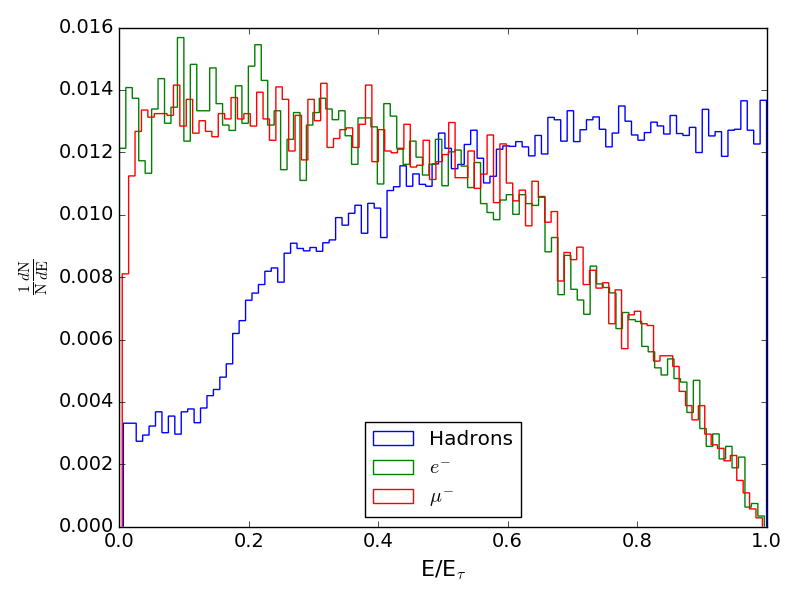}
\caption{Fractional energy distributions of hadronic and leptonic daughters from 100,000 tau decays generated with Pythia. The hadronic channel includes the sum of all hadrons in a single decay. The given distributions are for a relativistic tau with negative polarization (as the tau must carry the polarization of the parent neutrino), as this is the relevant case for any air shower study. The average fractional energy carried by hadrons is $58\%$. For electrons, and muons, this value is $41\%$.}
\label{Daughter_Distributions}
\end{wrapfigure}

For the experiments seeking to measure UAS, event estimations usually assume shower development similar to proton or gamma induced air showers with some average fractional energy deposition into decay products (roughly $50\%$ of the primary tau lepton energy), taking into account the decay length of the tau in the atmosphere. This simplification is useful for two reasons. 1) The tau lepton has multiple decay branches, of which $65\%$ are hadronic, and an additional $18\%$ are electromagnetic and 2) The physics of proton and gamma induced cosmic ray air showers are fairly well known and well parameterized, due in large part to their frequent study in collider and air shower experiments. However, this begs the question: what information is lost in approximating tau showers as "decay delayed" proton showers and does it strongly affect the number of possible neutrino events that can be recorded? The tau decay has a rich phenomonology. Additionally, the products resulting from a tau lepton decay do not have trivial energy distributions, nor are they distributed identically. The fractional energy distributions of the hadronic channel, and leptonic channels of the tau decay are shown in figure \ref{Daughter_Distributions} as calculated with Pythia simulations \cite{pythia}.


As of yet, there has not been a definitive effort to parameterize high energy tau lepton induced air showers. This work serves as a first step to quantify how tau lepton induced air showers are different from those induced by conventional primaries and what (if any) implications there may be for future tau neutrino observatories.

\section{Comparison to Conventional Primaries}
To simulate tau lepton air showers, we use a modified version of CORSIKA-75600 \cite{corsika}, which allows us initiation of showers at true ground level. The geometry we use for our simulations is upgoing, and nearly perfectly horizontal (to allow for a sufficient upper bound on the total slant depth to use in longitudinal profiles). We have also modified the decay times of tau leptons in CORSIKA such that they decay immediately. In this manner, we monitor the shower induced by the tau decay products without the possibility of decay outside the atmosphere. This is also to say that we ignore the ionization energy losses of the tau lepton through air, as it is a very small effect. Later, we adjust the longitudinal profiles by defining a tau decay point, distributed exponentially with mean $c \tau_{\mathrm{lifetime}} E_{\tau}/m_{\tau} \approx 5~\mathrm{km}(E_{\tau}/10^{17}~\mathrm{eV})$. The atmospheric model we use is that of the US Standard Atmosphere.

We simulate 1000 tau lepton events for energies between 1~PeV to 10~EeV, spaced equally by 1 in log space, and for each shower, record the 4 Gaisser-Hillas fit parameters ($\mathrm{N}_{\mathrm{max}}$--number of charged particles at the shower maximum, $X_{\mathrm{max}}$--atmospheric depth of the shower maximum, $X_{0}$--shower shaping parameter commonly thought of as the depth of first interaction, and $\Lambda$--shower shaping parameter which determines thickness and asymmetry of the profile) of the longitudinal charged particle profile using a least squares fit \cite{Gaisser}. Note that in a high energy tau decay, which may initiate deep into the atmosphere, the only parameters to be affected are $X_{\mathrm{max}}$ and $X_{0}$, which will leave the width and the shape of the shower intact. We then do the same for gamma and proton primaries in the same energy ranges.

To make a fair comparison against the tau intiated showers, we perform a sampling process on the proton and gamma initiated showers. We sample a fractional energy of the tau decay which goes into showering products (hadrons and electrons) from the Pythia simulations shown in figure \ref{Daughter_Distributions} and multiply by the primary energy of interest. From this sampled energy, we calculate the different shower parameters from proton and gamma showers using the data we previously generated. For each primary energy, we sample 1000 different events.

\begin{figure}
\begin{tabular}{cc}
  \includegraphics[width=.5 \textwidth]{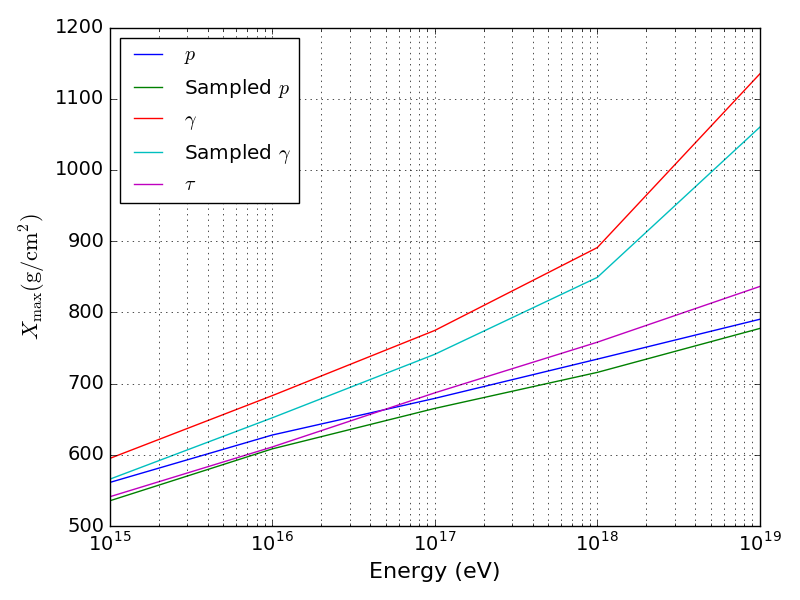} & 
  \includegraphics[width=.5 \textwidth]{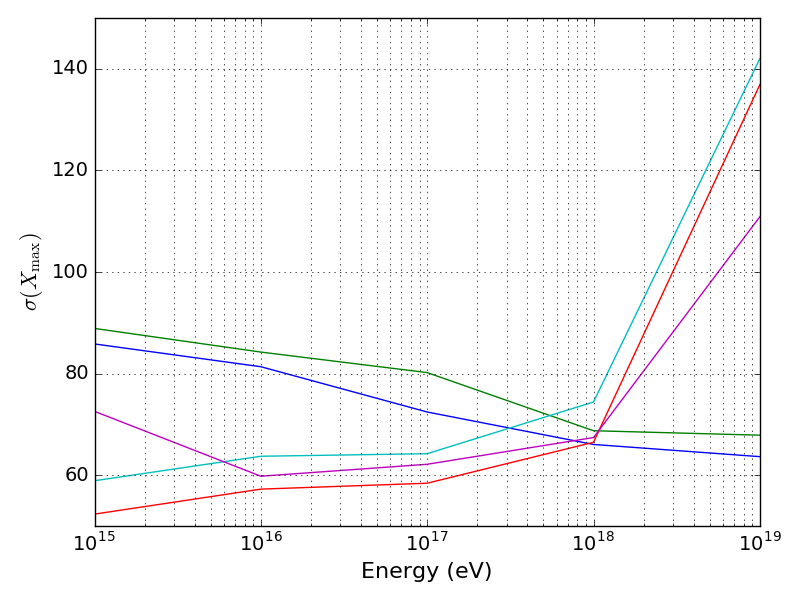}\\
(a) & (b) \\[6pt]
  \includegraphics[width=.5 \textwidth]{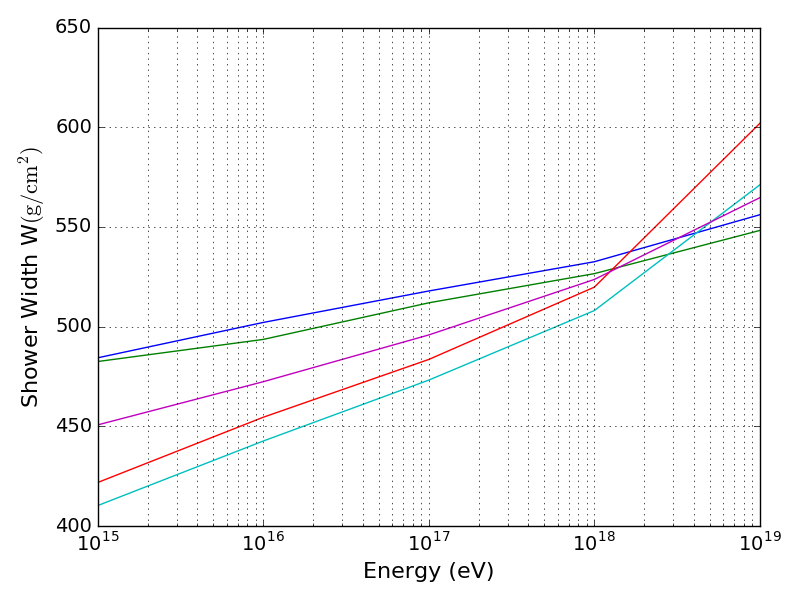} & 
  \includegraphics[width=.5 \textwidth]{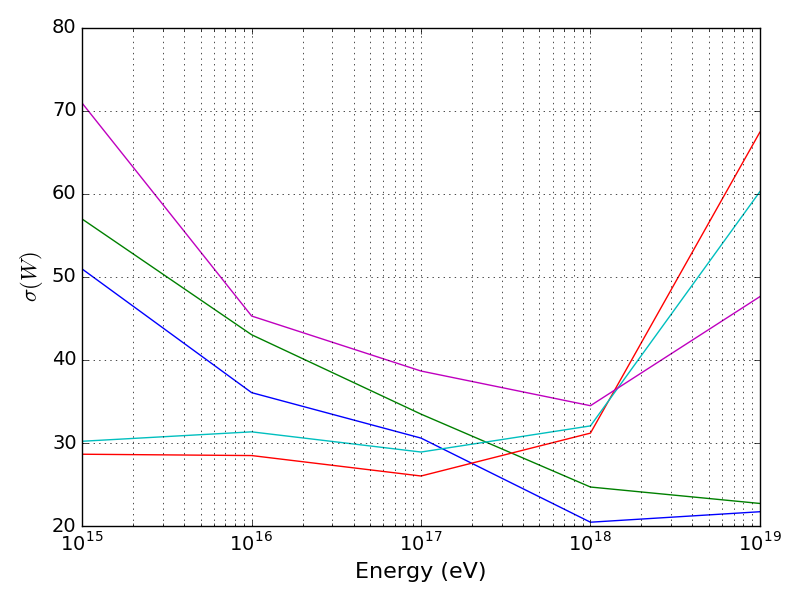}\\
(c) & (d) \\[6pt]
  \includegraphics[width=.5 \textwidth]{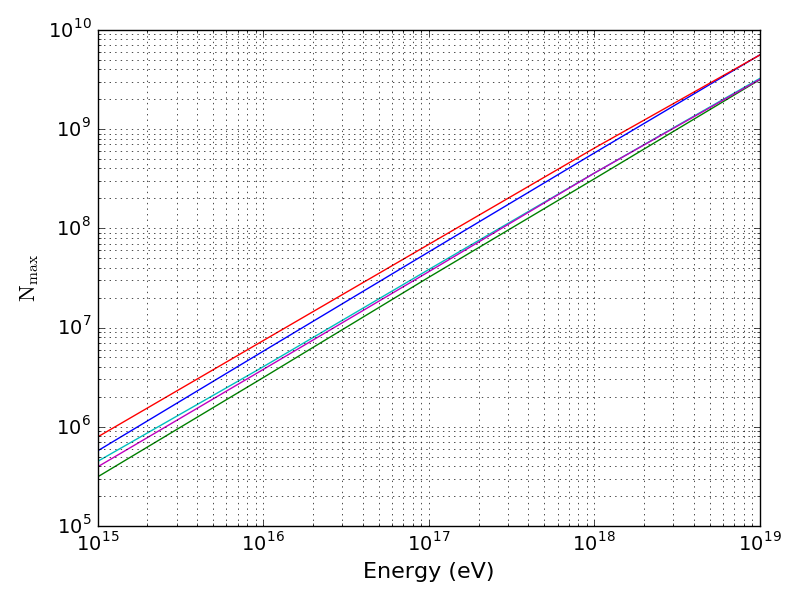} & 
  \includegraphics[width=.5 \textwidth]{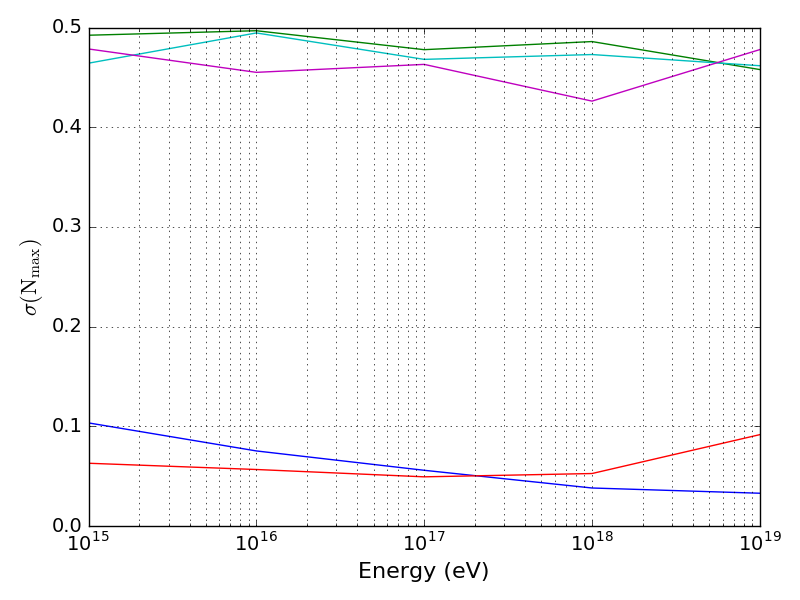}\\
(e) & (f) \\[6pt]

\end{tabular}
 \caption{a) $X_{\mathrm{max}}$ b) $\sigma(X_{\mathrm{max}})$ c) Shower Width W d) $\sigma(W)$ e) $\mathrm{N}_{\mathrm{max}}$ and f) $\sigma(\mathrm{N}_{\mathrm{max}}$) as functions of energy for showers initiated by tau leptons, protons, gammas, and protons and gammas subject to our sampling process described in the text.} 
\label{Tau_Comp}
\end{figure}

In figure \ref{Tau_Comp}, we plot as a function of energy the average $X_{\mathrm{max}}$, shower width W, and $\mathrm{N}_{\mathrm{max}}$ as well as RMS values for showers initiated by tau leptons, protons, gammas, and sampled protons and gammas. W is calculated via $2 \sigma \sqrt{2 \mathrm{ln} 2} \left[ 1+\frac{\mathrm{ln} 2}{18} z^{2}\right]$, where $z = \sqrt{\Lambda/(X_{\mathrm{max}}-X_{0})}$  and $\sigma =\sqrt{\Lambda (X_{\mathrm{max}}-X_{0})} $\cite{Lipari}. Excluding those points which come from muon induced showers via external cuts (this will be detailed in the following section) and across nearly all energies, the average $X_{\mathrm{max}}$ of a tau shower is slightly larger than that of proton showers (and significantly less than that of gamma initiated showers) of comparable energy. Similarly, the average shower width of tau initiated showers is also, on average, lower than that of proton showers and higher than that of gamma initiated showers. The average $\mathrm{N}_{\mathrm{max}}$ for a tau lepton shower is very well approximated by taking $50\%$ of $\mathrm{N}_{\mathrm{max}}$ of either proton or gamma initiated showers, as many estimations do.

Tau lepton induced air showers will deposit roughly the same energy into cascading particles as proton and gamma induced air showers of comparable energy. However, they will deposit their energy, on average, deeper in the atmosphere than proton showers, and shallower than gamma showers. Similarly, for energies under $10^{18}$~eV, tau showers deposit energy over a shorter path length than proton showers, and longer path length than gamma showers. This behavior changes at higher energies due to the LPM effect on high energy gammas \cite{Heck}. Because the tau lepton decays electromagnetically and hadronically (with many electromagnetic showers initiated also by the decay of the $\pi^{0}$ to $2\gamma$), the most accurate shower parameterization for tau leptons will be a weighted average of the parameterizations for proton and gamma showers. However, we note that if one wants to approximate with only one species, protons are, in general, a better approximation than gammas.

\section{Muons}
Muons are an interesting component of the tau shower phenomenology that is usually overlooked, despite taking up nearly 20 percent of the tau decay spectrum. Because of their comparably low interaction cross sections, it is assumed that they will not interact in the atmosphere in a noticeable way. An estimation using standardized muon cross sections demonstrates that this is not such a straightforward conclusion, especially for experiments which observe large portions of the atmosphere, as balloon and space based instruments intend to do.

\begin{figure}[t!]
\begin{tabular}{cc}
  \includegraphics[width=.5 \textwidth]{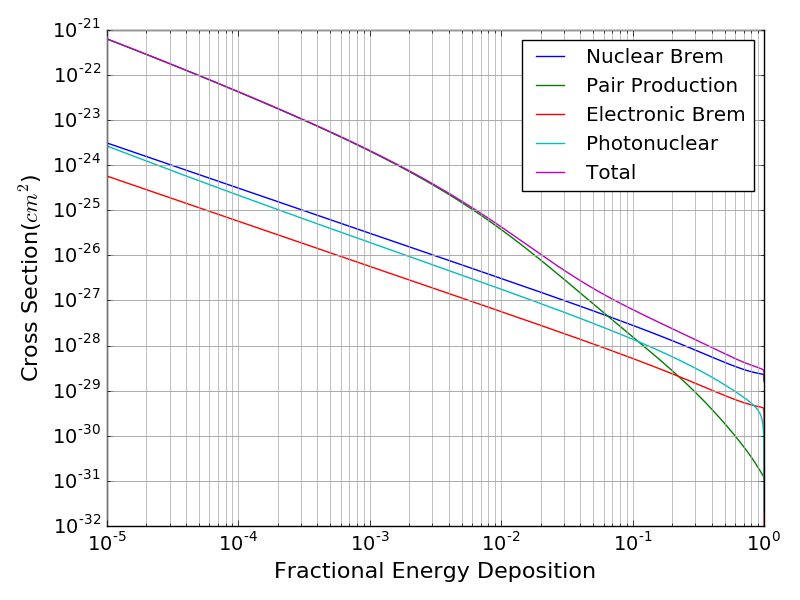} & 
  \includegraphics[width=.5 \textwidth]{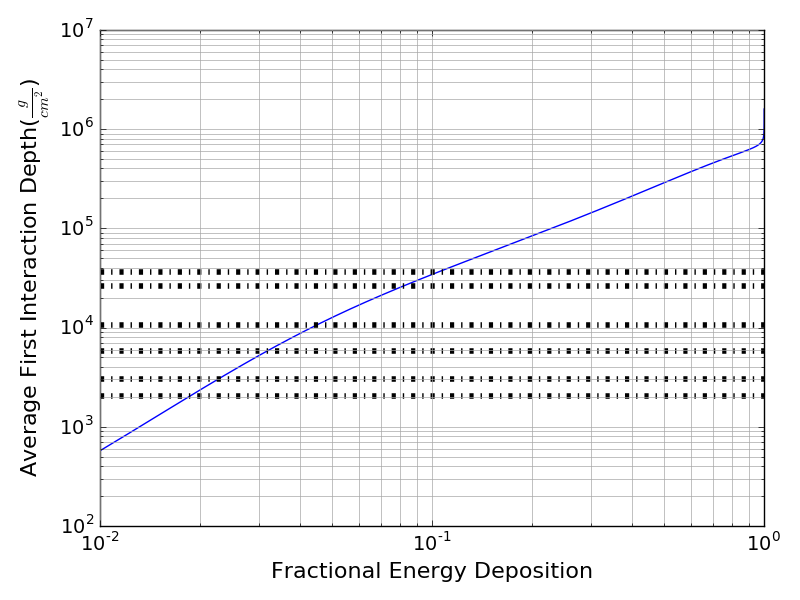}

\end{tabular}
 \caption{ Left: Muonic differential cross sections for nuclear and electronic brehmsstrahlung, photonuclear interactions, and electron positron pair production as a function of fractional energy deposition for a 1~PeV muon in air. Right: Corresponding average interaction depth in Earth's atmosphere. Dashed lines correspond to the full thickness of Earth's atmosphere for emergence angles $30^{\circ}$, $20^{\circ}$, $10^{\circ}$, $5^{\circ}$, $1^{\circ}$, and $0^{\circ}$ (perfectly horizontal). Brehmsstrahlung and Photonuclear cross sections are taken from \cite{muon1}, whereas pair production is more conveniently described in \cite{muon2}.}
\label{Muon_Cross_Sections}
\end{figure}

For a first approximation, we shall only consider processes which may deposit large amounts of the primary muon energy in a single interaction (nuclear and electronic bremsstrahlung emission, photonuclear interactions, and electron positron pair production), as they are the most relevant processes for energies above 1~PeV and high fractional energy depositions (which may then develop into a shower) \cite{muon1} \cite{muon2}. Additionally, we will ignore the possibility of muon decay, as the decay length of a muon is $6.25*10^{6} \Big(\frac{E}{1~\mathrm{PeV}} \Big)$~km, whereas the maximum path length through Earth's atmosphere is roughly 1200~km. The differential muon cross sections are shown in figure \ref{Muon_Cross_Sections}.

Above 1PeV, the nuclear bremsstrahlung cross section is nearly independent of muon energy, whereas the electronic brehmsstrahlung and pair production cross sections increase as the log of energy, and the photonuclear cross section increases as the log of energy squared. Across our energy range, this will lead to slight increases of $25\%$ for electronic brehmsstrahlung and pair production and $60\%$ for photonuclear, which we will ignore for brevity here, but indicate that for higher energy muons the probability to interact is even higher than what is listed here. This is to say that our values included here should be considered a lower bound on muon shower probabilities. To verify that our cross sections well describe reality, we simulate 1000 1~PeV muon showers in CORSIKA and calculate the cumulative probability that a muon deposits at least a fractional energy $v$ in a full atmosphere, and then compare to the analytical solution via:

\begin{equation}
\mathrm{P}_{\mathrm{int}} (X) = 1-e^{-X \sigma N} \quad\text{where}\quad 
\sigma = \int_{v}^{1} \frac{d \sigma}{d v} dv
\end{equation}

\begin{figure}[t!]
\begin{tabular}{cc}
  \includegraphics[width=.5 \textwidth]{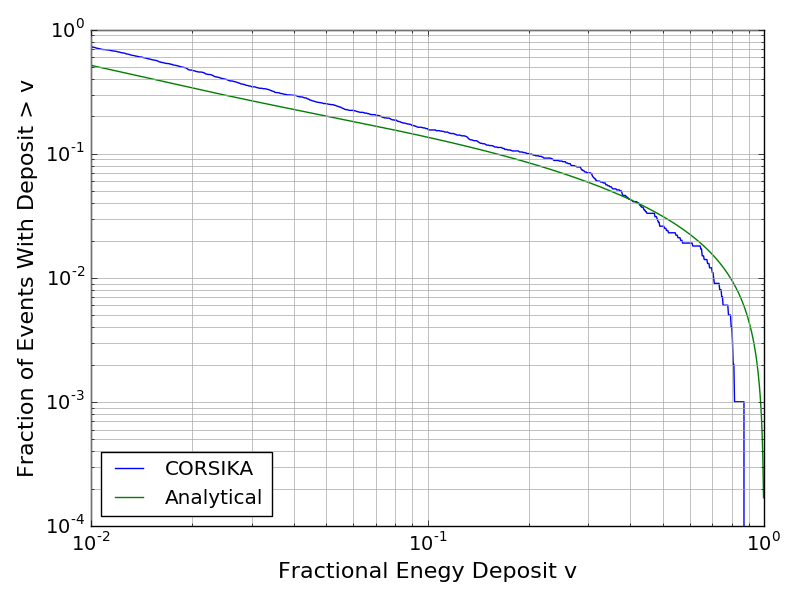} & 
  \includegraphics[width=.5 \textwidth]{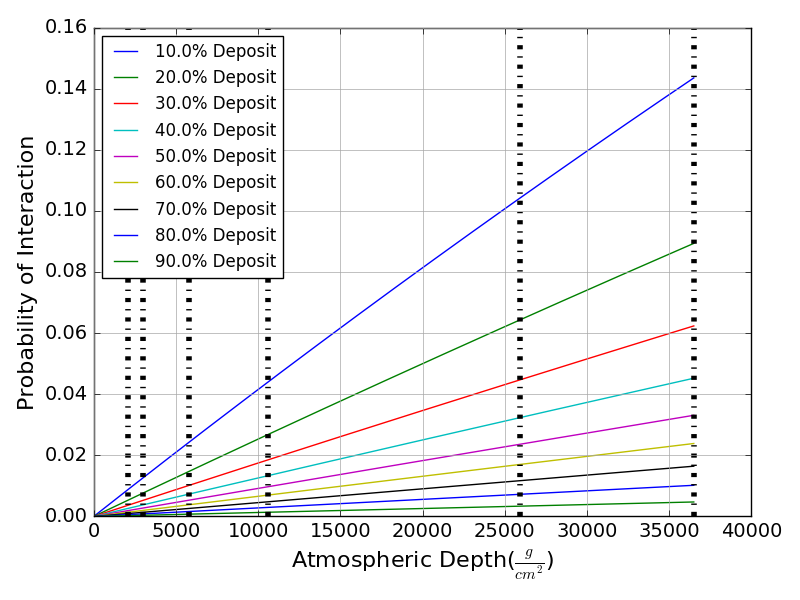}

\end{tabular}
 \caption{Left: Cumulative interaction probability for a full atmosphere ($X = 35000 \mathrm{g}/\mathrm{cm}^{2}$) calculated from CORSIKA showers and muon cross sections. Right: Muon interaction probability as a function of atmospheric depth for various fractional energy distributions. Vertical lines are the same as those in figure \ref{Muon_Cross_Sections}}
\label{Full_Atmos_Prob}
\end{figure}

With $N$ being the number of targets in one gram of air. The cumulative interaction probability as a function of fractional energy deposit calculated from these CORSIKA showers and from the muon cross sections is demonstrated in figure \ref{Full_Atmos_Prob}a, which shows good agreement. The cumulative interaction probability as a function of depth is shown in figure \ref{Full_Atmos_Prob}b for various fractional energy depositions.




Figure \ref{Full_Atmos_Prob} shows that a non-negligible percentage of muons experience large energy losses inside Earth's atmosphere, which begins a conventional particle cascade, with over $14\%$ of muons depositing more than $10\%$ of their energy inside a full atmosphere, to state one numeric example. Therefore, a flux of high energy tau leptons exiting the Earth from neutrino interactions guarantees a flux of high energy muons, which have a non-negligible chance to develop comparably strong particle cascades initiated at significantly greater depths in the atmosphere.


This allows us to draw some preliminary conclusions as to which experimental designs this muon component may be a relevant background or even a detectable signal. For muons to develop observable showers, very large amounts of atmosphere are necessary, independent of initial muon energy. Experiments which aim to observe tau neutrinos  with energies greater than $10^{17}$~eV through the UAS technique must necessarily observe large path lengths through the atmosphere so as not to exclude tau leptons with long decay lengths. The balloon based EUSO-SPB2, and satellite based POEMMA, will view the full Earth atmosphere at the limb, for instance \cite{JFK}. This makes ultra-high energy tau neutrino observatories an ideal candidate to observe also showers induced by muons. For instance, we note that in the Cherenkov emission channel, for a specific angular and energy range, this signal from the muonic decay channel of the tau could be stronger than that of showers initiated by the hadronic (and electronic) decay channel \cite{me}.

\subsection{Cherenkov Emission from Tau Lepton Showers}
\begin{figure}[t!]
\begin{tabular}{cc}
  \includegraphics[width=.5 \textwidth]{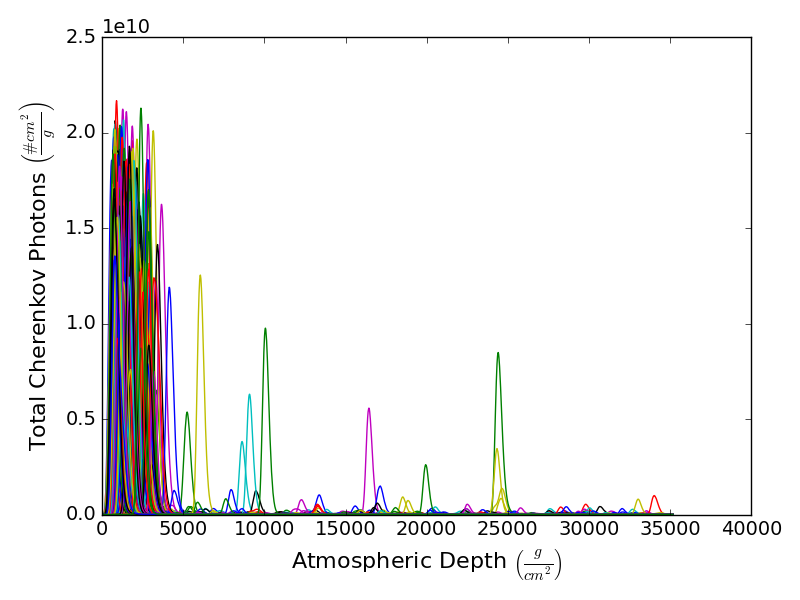} & 
  \includegraphics[width=.5 \textwidth]{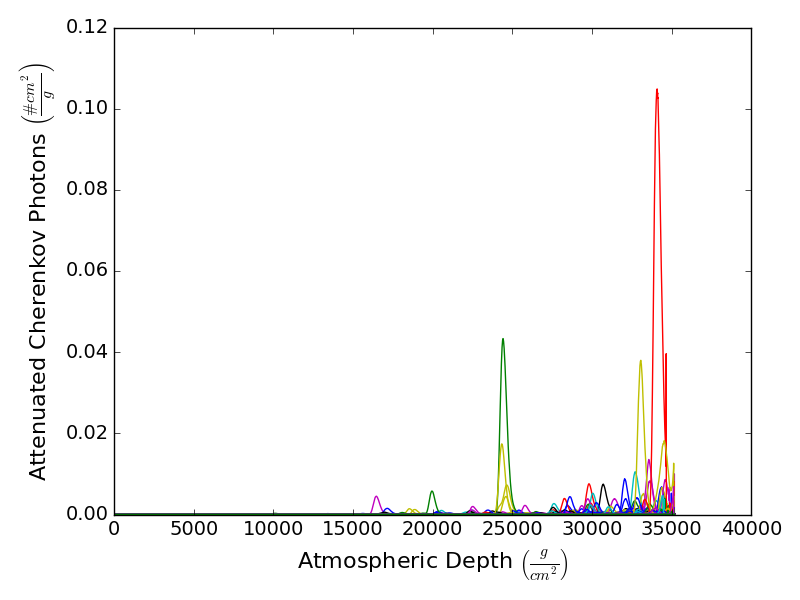}

\end{tabular}
 \caption{Left: Total generated Cherenkov photons and Right: Cherenkov photons including atmospheric attenuation and geometric correction factors (for an instrument at 525~km altitude) from 1000 100~PeV tau lepton showers with 0.1 degree Earth emergence angle in CORSIKA. Observe that even muons from tau decay which deposit low amounts of energy still appear significantly brighter than showers initiated by the hadronic decay branch of the lepton.}
\label{Cherenkov_From_Tau}
\end{figure}

Atmospheric extinction of visible light is remarkably strong for wavelengths below 450~nm, where Cherenkov emission is relevant. Thus, it is difficult to measure any Cherenkov photons resulting from charged particles in the lower atmosphere (exponentially moreso for low inclination angles) \cite{Atmosphere}. However, the first interaction point of muons can be of the order of the total thickness of the atmosphere, so any Cherenkov emission from the shower they produce is subject to significantly less attenuation (compared to showers initiated by hadronic or electronic decay products), as well as a more focused Cherenkov cone and a shorter distance between point of creation and point of detection. All of these effects aid in producing a significantly higher photon flux for a high altitude instrument. Cherenkov profiles of tau leptons both with and without atmospheric attenuation and geometric correction for a space based instrument are shown in figure \ref{Cherenkov_From_Tau} to help illustrate this effect.

We have briefly explained here the idea that high energy muons may be detectable by a space or balloon based instrument via Cherenkov emission from a charged particle shower, and for a certain angular range, can be even brighter than the corresponding showers from the hadronic decay of the tau lepton. We have begun work to show the effect of including the muon decay branch of tau decays has on conventional estimates of tau neutrino sensitivity. Additionally, we have begun work to explore the viability of measuring the muon neutrino flux using the UAS technique, noting that the severe energy losses of the muon in the Earth may be counterbalanced by the improved signal quality \cite{me}. This is beyond the scope of these proceedings, but it is an important conclusion to state which would not have been realized without analyzing the tau in detail.

\section{Summary}
In this work, we simulated thousands of upward tau lepton showers in CORSIKA for energies between $10^{15}$~eV to $10^{19}$~eV and compared them to showers initiated by proton and gamma primaries using the Gaisser-Hillas parameterization \cite{Gaisser}. We found that approximation of the hadronic decay channel of the tau as a proton or gamma primary of sampled equivalent energy is reliable and reproducible, and estimations based on this assumption are also likely safe from significant scrutiny. Although, we do note that, in general, it is more accurate to use a proton parameterization than a gamma parameterization for these purposes.

Additionally, we analyzed the muon decay channel of the tau, and determined that it is not fair to treat it as a negligible signal as it often is, especially for experiments which seek to observe tau neutrinos with energies greater than $10^{17}$~eV. We illustrated that not only do a significant number of muons conventionanlly shower in large amounts of atmosphere, but that these showers may often appear as stronger signals than their hadronic counterparts. From this, we have begun work to show the effect this has on tau neutrino detection via the UAS approach, and to explore whether muon neutrinos can be detected in the same manner \cite{me}.

\end{document}